\documentclass[conference,final]{IEEEtran}

\usepackage[top=0.65in, bottom=0.94in, left=0.6in, right=0.6in]{geometry}

\usepackage{booktabs}
\usepackage{amsmath}
\usepackage{amsfonts}
\usepackage{amssymb}
\usepackage[pdftex]{graphicx}
\usepackage{listings}
\usepackage{bbding}
\usepackage{multicol}
\usepackage[ruled,vlined]{algorithm2e}

\usepackage{array}
\newcommand{\PreserveBackslash}[1]{\let\temp=\\#1\let\\=\temp}
\newcolumntype{C}[1]{>{\PreserveBackslash\centering}p{#1}}
\newcolumntype{R}[1]{>{\PreserveBackslash\raggedleft}p{#1}}
\newcolumntype{L}[1]{>{\PreserveBackslash\raggedright}p{#1}}

\newcommand\bsymbdefop[2]{
  \newcommand{#1}{\mathop{#2}\nolimits}
}
\newcommand\bsymbdeford[2]{
  \newcommand{#1}{\mathord{#2}}
}
\bsymbdeford{\bfalse}{\bot}
\bsymbdeford{\btrue}{\top}
\newcommand{\limp}{\mathbin\Rightarrow}

\bsymbdeford{\qdot}{\mkern1mu\cdot\mkern1mu}
\bsymbdefop{\partition}{\mathrm{partition}}
\newcommand\defi{\mathrel{\widehat=}}

\bsymbdefop{\pow}{\mathbb P\hbox{}}
\bsymbdefop{\pown}{\mathbb P_1}
\newcommand{\cprod}{\mathbin\times}
\newcommand{\bunion}{\mathbin{\mkern1mu\cup\mkern1mu}}

\bsymbdefop{\union}{\mathrm{union}}
\bsymbdefop{\inter}{\mathrm{inter}}

\renewcommand{\emptyset}{\mathord\varnothing}

\bsymbdefop{\dom}{\mathrm{dom}}
\bsymbdefop{\ran}{\mathrm{ran}}

\bsymbdefop{\id}{\mathrm{id}}

\newcommand{\domsub}{\mathbin{\lhd\mkern-14mu-}}

\newcommand{\ovl}{\mathbin{\lhd\mkern-9mu-}}

\bsymbdefop{\prjone}{\mathrm{prj}_1}
\bsymbdefop{\prjtwo}{\mathrm{prj}_2}

\newcommand{\bsymbpartial}[2]{
  \mathbin{\mkern#2mu\mapstochar\mkern-#2mu#1}
}
\newcommand{\pfun}{\bsymbpartial\rightarrow6}
\newcommand{\tfun}{\mathbin\rightarrow}
\newcommand{\pinj}{\bsymbpartial\rightarrowtail9}

\bsymbdeford{\nat}{\mathbb N}
\bsymbdeford{\natn}{\mathbb N_1}
\bsymbdeford{\intg}{\mathbb Z}

\bsymbdefop{\finite}{\mathrm{finite}}
\bsymbdefop{\card}{\mathrm{card}}
\bsymbdefop{\upred}{\mathrm{pred}}
\bsymbdefop{\usucc}{\mathrm{succ}}

\bsymbdeford{\Bool}{\mathrm{BOOL}}
\bsymbdeford{\True}{\mathrm{TRUE}}
\bsymbdeford{\False}{\mathrm{FALSE}}
\bsymbdefop{\bool}{\mathrm{bool}}

\usepackage{color}
\definecolor{keycolor}{rgb}{0,0,0}  
\definecolor{labelcolor}{rgb}{0,0,0}  
\definecolor{codecolor}{rgb}{0,0,0}  
\definecolor{ccodecolor}{rgb}{0,0.2,0}  
\definecolor{cmtcolor}{rgb}{0,0,0}  
\definecolor{xcolor}{rgb}{1,0,0}  

\newcommand{\evtbmach}[1]{$\mathrm{#1}$}

\newcommand{\arinctext}[1]{$\mathrm{#1}$}

\newcommand{\sectprefix}{{Section}}
\newcommand{\subsectprefix}{{Subsection}}
\newcommand{\figprefix}{{Fig.}}
\newcommand{\algprefix}{{Algorithm}}
\newcommand{\tabprefix}{{Table}}

\begin{document}

\title{\LARGE\textbf{Event-based Formalization of Safety-critical Operating System Standards: An Experience Report on ARINC 653 using Event-B}}

\author{\IEEEauthorblockN{Yongwang Zhao$^{*,\dag}$, Zhibin Yang$^{\ddag}$, David San\'an$^{\dag}$ and Yang Liu$^{\dag}$}
\IEEEauthorblockA{\small \emph{$^{*}$School of Computer Science and Engineering, Beihang Univerisity, Beijing, China}\\
\emph{$^{\dag}$School of Computer Engineering, Nanyang Technological University, Singapore}\\
\emph{$^{\ddag}$College of Computer Science and Technology, Nanjing University of Aeronautics and Astronautics, Nanjing, China}\\
\emph{Email: zhaoyw@buaa.edu.cn}}
}

\maketitle

\begin{abstract}
	Standards play the key role in safety-critical systems. Errors in standards could mislead system developer's understanding and introduce bugs into system implementations. In this paper, we present an Event-B formalization and verification for the ARINC 653 standard, which provides a standardized interface between safety-critical real-time operating systems and application software, as well as a set of functionalities aimed to improve the safety and certification process of such safety-critical systems. The formalization is a complete model of ARINC 653, and provides a necessary foundation for the formal development and verification of ARINC 653 compliant operating systems and applications. Three hidden errors and three cases of incomplete specification were discovered from the verification using the Event-B formal reasoning approach.
\end{abstract}

\IEEEpeerreviewmaketitle

\section{Introduction}
\label{sec:intro}
In recent years, safety-critical systems have paved the way for the integration on one single platform of different criticality level application subsystems developed by different vendors, like the Integrated Modular Avionics (IMA) \cite{Rushby00} in avionics domain. IMA aims Partitioning Operating Systems (POS) implementation, supporting spatial and temporal partitioning~\cite{ARINC653p1}. 
Partitioning \cite{Rushby00} provides independent execution of one or more applications, which procures temporal and spatial separation and fault containment to prevent propagation of application failures. In this way, partitioning is equivalent to an idealized system in which each partition allocates an independent processor and associated peripherals for the execution of applications, where all inter-partition communications are carried out on dedicated lines, giving applications with an environment which is undistinguishable from that provided by a physically distributed system.

In avionics industry, ARINC 653 \cite{ARINC653p1} was first published in 1996 as a set of specifications to guide manufacturers in avionic application software towards maximum standardization. It aims to provide a standardized interface between a given POS and application software, as well as a set of functionalities to improve safety and certification process of safety-critical systems. ARINC 653 compliant POSs have been widely applied in safety-critical domains. 
Typical POSs are VxWorks 653 platform, INTEGRITY-178B, LynxOS-178, PikeOS, and open source software, e.g. POK \cite{Delange11} and Xtratum \cite{Masmano09}. 
The ARINC 653 standard can be considered as a requirement for POSs under construction and a base for compliance tests on their implementation. However, hidden inconsistencies or incorrectness in standards could mislead system developers' understanding, causing failures or malfunction in POSs, and hence a breakdown of applications, which is not allowed in safety critical systems. For this reason, ensuring standards correctness, and in particular ARINC 653 correctness, w.r.t. a set of functional, safety, and security properties, is necessary in ensuring absence of failures on safety-critical systems development.

Due to POSs' complexity, traditional test-based techniques are not enough to warrant their correctness, as it is not possible to generate all necessary test cases to fully cover all behaviours of POSs, and hence to ascertain their correctness. During last decades, formal methods based techniques have been widely applied on the verification of both software and hardware \cite{Woodcock09}. In the field of real-time operating systems for safety-critical systems, most of related work has been concentrated on specifying or verifying the operating system \cite{zhao15}. Nevertheless, as the interface of the fundamental execution environment of IMA applications, a formal model of ARINC 653 is strongly necessary for formal development and verification of ARINC 653 compliant operating systems and ARINC 653 based applications. Although some research efforts have been paid off in the formal modelling and verification of ARINC 653 based systems (\cite{de07,Singhoff07,Oliveira12,Wang11,Delan10}), to our knowledge the formalization introduced in this work is the most complete model of the ARINC 653. We provide a detailed comparison with related work at {\sectprefix} \ref{sec:bg}.

This paper presents the formalization of ARINC 653 Part 1 (the latest version, Version 3) \cite{ARINC653p1}, its verification using a deductive verification approach \cite{Beck14}, and the errors in the standard found out during the verification. An ARINC 653 formal model is constructed using Event-B \cite{Abrial07}, a mathematical approach based on a model-driven design methodology used for specifying and reasoning about complex systems, including concurrent and reactive systems. Event-B uses the set theory as a modelling notation, refinement to represent systems at different abstraction levels, and mathematical proofs to verify consistency between refinement levels \cite{Abrial07}. We choose Event-B due to the following reasons: (1) a specification in Event-B is easy to understand and has a strong development environment-Rodin, for which there exists many plugins to translate Event-B specifications into other formalization and source code, and for model visualization and simulation, among other functionalities; (2) its high degree of automatic reasoning eases the verification, and the inductive approach avoids state space explosion when verifying complicated systems; (3) \emph{events} are very suitable for modelling operating systems, where hardware components, e.g. interrupters like clock and timers, need to be well managed.

We have modelled the system functionality of POS and all of 57 services specified in ARINC 653 Part 1, including partition and process management, time management, inter- and intra- partition communication, and health monitoring. 
We use the deductive verification approach supported in Event-B and its Rodin platform \cite{rodin} as a means for consistency checking. The description of the \emph{system functionality} in ARINC 653 is manually formalized as the top level specification and safety properties (invariants), and the \emph{service requirements} are translated into the low level specifications. Safety properties on the specifications and refinement between the top level and low level specifications are proven by discharging proof obligations. Finally, we found three errors in ARINC 653 Part 1, amongst which, one in service of process management, and two in services of inter- and intra-partition communication. Additionally we detected three cases where the specification of process state transitions is incomplete.


The rest of this paper is organized as follows. In {\sectprefix} \ref{sec:bg}, we introduce the background and related work. Our framework approach is presented in {\sectprefix} \ref{sec:approach}. The formalization of ARINC 653 is presented in {\sectprefix} \ref{sec:form}. {\sectprefix} \ref{sec:result} presents the formalization and verification result, and a discussion. Finally, {\sectprefix} \ref{sec:concl} gives the conclusion and future work.

\section{Background and related work}
\label{sec:bg}

\subsection{Event-B}
\label{subsec:event-b}
Here, we provide a brief overview of Event-B. Full details are provided in \cite{Abrial13}. 
The Event-B method \cite{Abrial07} is used to build reliably systems using \emph{discrete system models} and aims at obtaining systems which can be considered to be \emph{correct by construction}, in the sense that the systems produced are guaranteed to implement the corresponding functional specification. 


Event-B models are described in terms of \emph{contexts} and \emph{machines}. Contexts specify the static part of a model whereas machines specify the dynamic part. 
Suppose a machine $M$, seeing a context $C$ with sets $s$ and constants $c$. An event of this machine is represented as

\begin{equation}
\begin{aligned}
E\; \defi \; & \mathbf{any} \; x \; \mathbf{where} \; G(s,c,v,x) \; \\ 
&\mathbf{then} \; v :\mid BA(s,c,v,x,v') \; \mathbf{end}
\end{aligned}
\end{equation}

$E$ is the event name, $G(s,c,v,x)$ is the \emph{guard} of the event that states the necessary condition for the event to occur, and $v :\mid BA(s,c,v,x,v')$ is the \emph{action} that defines how the state variables evolve when the event occurs. Actions use a \emph{before-after predicate}, which relates the values $v$ (before the action) and $v'$ (afterwards). 
In a machine, the execution of an event is considered to take \emph{no time} and no two events can occur simultaneously. When the guards of one or more events are true, one of these events necessarily occurs and the state is modified accordingly. Then, the guards are checked again, and so on.

Refinement in Event-B provides a means for introducing details about the dynamic properties of a model. A machine $RM$ can refine another machine $M$ and we call $M$ to the abstract machine (or refined machine), which specifies $RM$, and $RM$ the concrete machine (or refinement machine), which implements $M$. 


Event-B defines \emph{proof obligations}, which must be proven to show that machines hold the properties specified over them. 
For space reasons, we only describe here some of the most significant proof obligations. Formal definitions of all proof obligations are given in \cite{Abrial13}.
(1) \emph{Invariant preservation} states that invariants are maintained whenever variables change their values. It ensures events preserves invariants specified over a machine. 
(2) \emph{Guard strengthening} makes sure that the concrete guards in a concrete event are stronger than the abstract ones in the abstract event. This ensures that when a concrete event is enabled, so is the corresponding abstract one. 
(3) \emph{Simulation} makes sure that each action in an abstract event is correctly simulated in the corresponding refinement. This ensures that when a concrete event is executed its actions are not contradictory with the actions in the corresponding abstract event. 

Finally, the Rodin tool \cite{rodin} is an Eclipse-based IDE supporting the application of the Event-B method. This is an industrial-strength tool for creating and analyzing Event-B models. It includes a proof-obligation generator and support for interactive and automated theorem proving.

\subsection{ARINC 653}
\begin{figure}
\begin{center}
\includegraphics[width=2.2in]{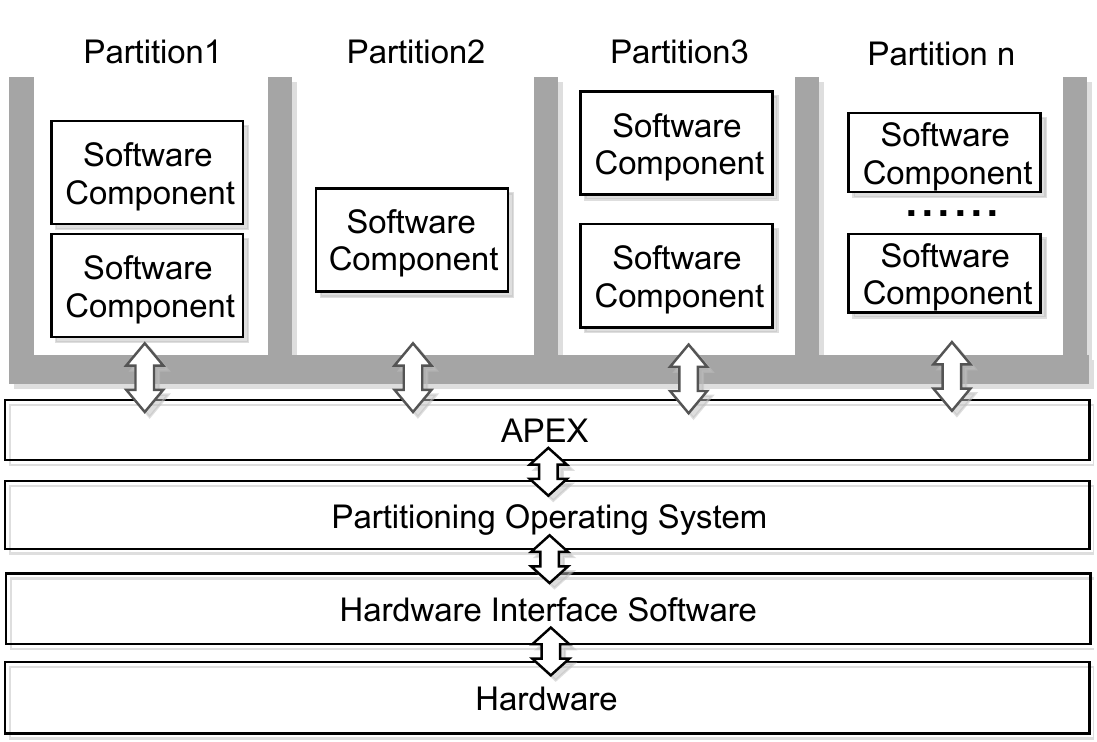}
\caption{System architecture based on partitioning operating systems} 
\label{fig:pos_arch}
\end{center}
\vspace{-10pt}
\end{figure}

The current version of ARINC 653 defines a partitioning architecture for safety-critical systems on single-core as shown in {\figprefix} \ref{fig:pos_arch}. A POS is in fact a small partitioning kernel that provides operating system services according to the safety features required by the safety integrity level.
The latest version of ARINC 653 published in 2010 is organized in six parts. Part 1, currently in Version 3, defines the standard APplication EXecutive (APEX) interface between the application software and the POS, and the list of services which allow the application software to control the scheduling, communication, and status information of its internal processing elements \cite{ARINC653p1}. We focus our work  in formalizing and verifying Part 1, since it specifies the baseline operating environment for application software used within IMA, and most of industrial and open source  POSs implementing ARINC 653 are compliant with this part. ARINC 653 Part 1 concentrates on specifying the \emph{system functionality}, which is described in natural language, and \emph{service requirements}, which is presented by a type of pseudo-code: the \emph{APEX service specification grammar}.

The required services specified in ARINC 653 Part 1 are grouped into the following major categories: partition management, defining partitions services, attributes, and the partition operating modes, the set of states a partition can be in and the transitions among them; process management, defining processes services, attributes, and the process operating modes; time management, defining time services for partitions and attributes; inter-partition communication, defining communication modes between partitions, services provided, and attributes; intra-partition communication, similar to inter-partition communication, but oriented to processes instead of partitions; and health monitoring, which defines actuation rules under system, partition, and application failures. 
Note that since partitions, and therefore their associated memory spaces, are defined during system configuration and initialization, there is no memory allocation service defined in APEX.
All required services are specified in detail by pseudo-code in the service specification grammar. For instance, the \emph{STOP} service from process management is illustrated in {\figprefix} \ref{fig:stop_service}.  For a detailed description of partitions, processes, and services provided by ARINC 653 we refer the reader to the ARINC 653 Standard~\cite{ARINC653p1}.

\begin{figure}
\begin{center}
\begin{lstlisting}[frame=single, breaklines=true, basicstyle=\tiny, numberstyle=\scriptsize] %\tiny,\scriptsize,\footnotesize,\small,

procedure STOP 
    (PROCESS_ID : in PROCESS_ID_TYPE; 
    RETURN_CODE : out RETURN_CODE_TYPE) is 
    
error 
    when (PROCESS_ID does not identify an existing process or identifies the current process) => 
        RETURN_CODE := INVALID_PARAM; 
    when (the state of the specified process is DORMANT) => 
        RETURN_CODE := NO_ACTION; 
        
normal 
    set the specified process state to DORMANT; 
    if (current process is error handler and PROCESS_ID is process which the error handler preempted) then 
        reset the partition's LOCK_LEVEL counter (i.e., enable preemption); 
    end if; 
    if (specified process is waiting in a process queue) then 
        remove the process from the process queue;
    end if;
    stop any time counters associated with the specified process;
    RETURN_CODE := NO_ERROR; 
end STOP; 
\end{lstlisting}
\caption{The service requirement of STOP service in process management} 
\label{fig:stop_service}
\end{center}
\vspace{-10pt}
\end{figure}


\subsection{Related work}
\label{subsec:rw}
A formal specification of the ARINC 653 architecture using the \textsf{Circus} language is presented in~\cite{Oliveira12}. 
Its specification focuses on the whole ARINC 653 architecture and interactions between IMA components.  However, the formal model only covers a small part of ARINC 653 services and no verification is carried out. 
Also focussing only in modelling, ARINC 653 components and their constraints are modelled using AADL (Architecture Analysis and Design Language) that can be used for model-driven development of IMA application (\cite{Wang11,Delan10}), but not all APEX services are covered in these works. In \cite{Singhoff07}, only an ARINC 653 hierarchical scheduler is modelled with AADL. 
Works in~\cite{de07,de11} target not only the ARINC specification but also its verification, where ARINC 653 services are modelled in PROMELA and verified used the SPIN model checker to ensure the correctness of avionics software constructed on top of ARINC 653. Here, the ARINC and the application models, which are extracted from the application's C source code, comprise the complete formal model for verification. However, the verification is focused only on process and time management, not covering any other ARINC 653 service or functionality.

The specification and verification of RTOSs and separation kernels are also related to our work,  as we overview in our technical report \cite{zhao15}.
It is worth noting that formal methods have been largely used in the specification of separation kernels, a generalization of POSs, like \cite{Verb14,Klaus15}, separation-partitioning micro-kernels~\cite{Sanan14}, or OSEK/VDX (\cite{Choi13,Vu12}), an international standard for automotive operating systems. 
In general, formal verification has been used on RTOSs for safety/security certification in industry, like in the AAMP7G microprocessor, which is a hardware implementation of partitioning in Rockwell Collins \cite{Wilding10}; in PikeOS, which is an ARINC 653 compliant POS in SYSGO AG (\cite{Baumann11,Tverdy11}); in INTEGRITY-178B which is also an ARINC 653 compliant POS in Green Hills \cite{Richards10}; and in an Embedded Devices kernel in Naval Research Laboratory \cite{Heitmeyer08}. It is worth to highlight the work in~\cite{Klein09}, where the sel4 microkernel has been fully verified from the top level specification down to the machine code.

The B-Method (predecessor of Event-B), has been also applied to operating systems. 
It has been used for the (partial) formal development of a secure partitioning kernel in the Critical Software company \cite{Andre09}. A real-time operating system, FreeRTOS, has also been formally specified using B method \cite{deharbe09}. The L4 microkernel, a kernel of general purpose OSs, has also been formally modelled in B (\cite{hoffmann06,kolan06}). Event-B extends B-Method with events, procedures that are activated when a guard is enabled, and it is suitable for modelling systems based on events. Both Event-B and B-Method share the same foundation, the main difference between them is that refinement in Event-B requires the guards of events of a concrete implementation to be stronger than the guards of the events in the abstract machine.

Formal verification on standards has attracted considerable attentions for a long time, such as communication standards/protocols (\cite{Devi00,Bharg02,Elank06,Eian12}). It has also been taken into account in safety-critical systems. Typical verification about standards in this domain is the verification of compliance to safety standards, such as in \cite{Pane11}. 


%
%

\section{Framework Approach}
\label{sec:approach}

We first present our framework approach in this section. 

\begin{figure}
\centering
\includegraphics[width=3.3in]{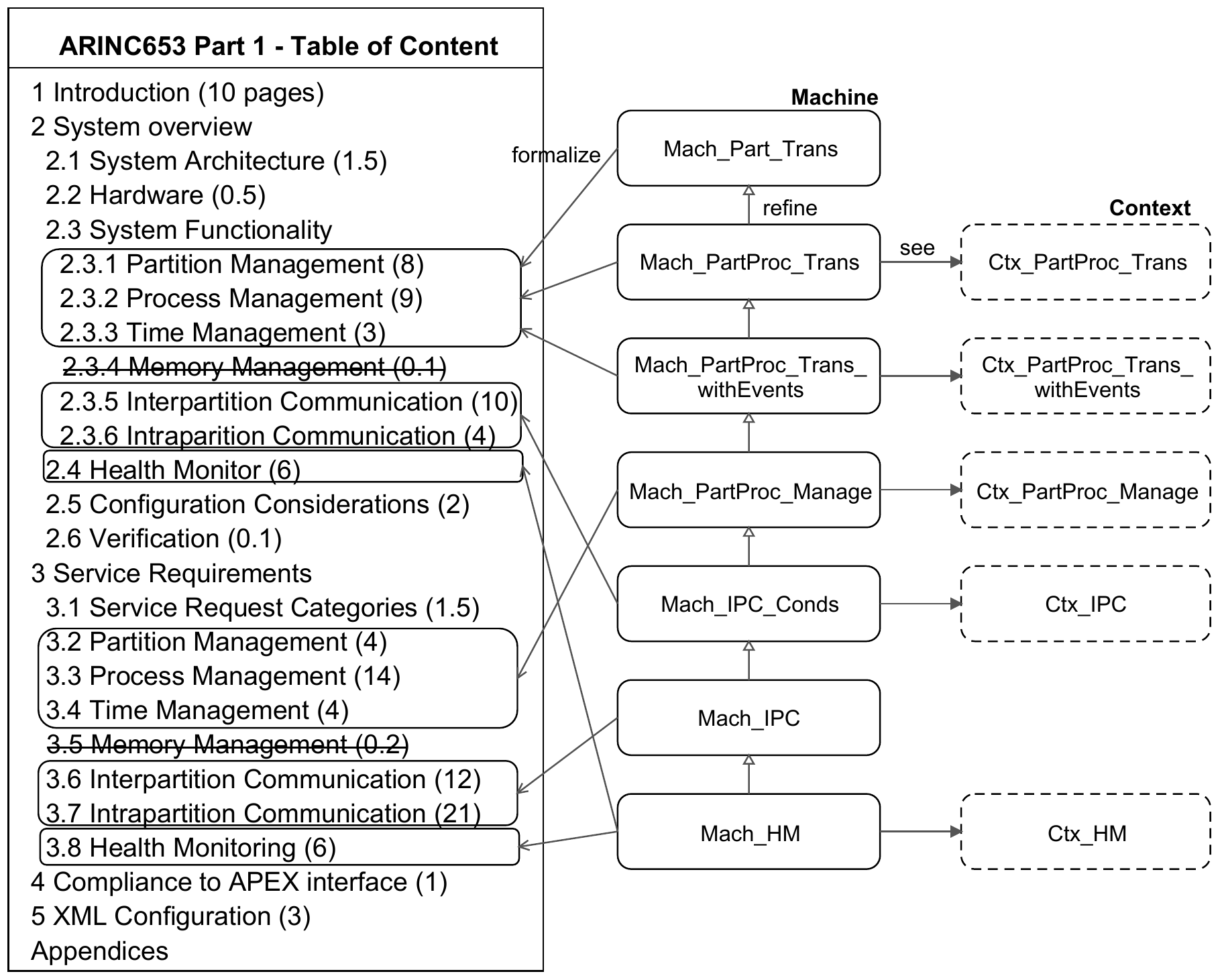}
\caption{ARINC 653 Part 1 and the Event-B model} 
\label{fig:model_overview}
\vspace{-10pt}
\end{figure}

\textbf{Modeling consideration}. The first aspect to be considered is \emph{what to be modelled}. The complete document structure and the number of pages of each section of ARINC 653 Part 1 are as shown in the left part of {\figprefix} \ref{fig:model_overview}. The content of ARINC 653 standard is divided into five parts: overview, system functionality, service requirements, configuration, and verification. Our work formalizes the system functionality (including \emph{health monitoring}) and service requirements, which are the main parts of the standard. Chapters 1, 2.1, 2.2 and 3.1 give an overview of ARINC 653 from different perspectives, which are a high level description for easy understanding and therefore need not to be modelled. The configurations described in 2.5 and 5 define the information and data format of the system configuration for integration and deployment, and are eliminated in our model. 
Notice that the chapter of \emph{memory management} in ARINC 653 does not define any service, and therefore is also eliminated in our model. Also, we omit other sections not mentioned here and not providing enough details/information to construct a formal model of them.

The second aspect into consideration is \emph{how to model}. The \emph{system functionality} is informally defined in natural language and can only be modelled manually. From this informal description, we extract components and their attributes, actions on components and their effects, and constraints. These elements are all represented in Event-B as \emph{sets}, \emph{constants}, \emph{variables}, \emph{events}, and \emph{invariants}. \emph{Service requirements} are presented by the \emph{APEX service specification grammar}, which is a structured language. It provides the possibility of semi-automatically translating these requirements to Event-B model. We design an algorithm to guide manual translation which is discussed in the next section.

\textbf{Model structure}. According to the document structure, we design our model formalization  as shown in the right part of {\figprefix} \ref{fig:model_overview}.
We firstly formalize the system functionality of partition, process, and time management. 
The Event-B machine \evtbmach{Mach\_Part\_Trans} models the partition operating modes. \evtbmach{Mach\_Part\_Proc\_Trans} refines the partition operating modes, and adds process state transitions. \evtbmach{Mach\_Part\_Proc\_Trans\_withEvents} defines all events of partition, process, and time management according to the system functionality. \evtbmach{Mach\_PartProc\_Manage} formalizes the service requirements of the partition, process, and time management. 
Then, we add the system functionality and service requirements of the communication: \evtbmach{Mach\_IPC\_Conds} specifies the functionality and events of inter-partition and intra-partition communication, and \evtbmach{Mach\_IPC} formalizes their service requirement. Finally, the system functionality and service requirements of  the health monitor are formalized in \evtbmach{Mach\_HM}.

\textbf{Verification}. The main verification approach in Event-B is deductive reasoning of proof obligations. \emph{Invariants} should be preserved on machines and refinement between a refined machine and its refinement (we mainly use \emph{guard strengthening} and \emph{simulation}). According to the model structure, the consistency of the system functionality and the service requirements is proven by refinements between \evtbmach{Mach\_Part\_Proc\_Trans\_withEvents} and \evtbmach{Mach\_PartProc\_Manage}, and \evtbmach{Mach\_IPC\_Conds} and \evtbmach{Mach\_IPC}. ARINC 653 defines the safety functionalities and variable's data type. Beside that, it does not explicitly define any property. The data type of each variable is defined as an \emph{invariant} on each variable in Event-B machines. We have also extracted all possible properties from the ARINC 653 standard, which are safety properties \cite{Schn87} as shown in {\tabprefix} \ref{tbl:invariants}. As explained in {\sectprefix} \ref{sec:result}, other extractable properties, such as \textit{liveness}, are not covered. 

From the informal description of ARINC 653, we use manual modelling of the system functionality and manual translation of the service requirements. Hence, we have to manually validate the errors found in our Event-B formalization. After finding an error in the Event-B model, we check corresponding description in the ARINC 653 standard to confirm and locate the error. 

\begin{table}
\centering
\scriptsize
\caption{Safety properties from ARINC 653} 
\begin{tabular}{p{0.1cm}p{8cm}} 
\toprule
No. &Functionality / Invariant description\\
\midrule
& \textbf{Partition and process management} \\
(1)& each process is in one partition \\
(2)& if a partition is not in $NORMAL$ mode, its processes should not in the state of $Ready$, or $Running$ or $Suspend$ \\
(3)& if there are processes of a partition in state of $Ready$, or $Running$ or $Suspend$, the partition's mode should be $NORMAL$  \\
(4)& if a partition's mode is $NORMAL$, it should have processes \\
(5)& if a partition's mode is $IDLE$, it should not have any process \\
(6)& there is at most one $Running$ process in a single core system \\
(7)& when a partition is in the $COLD\_START$ or $WARM\_START$ mode, the lock level should be larger than zero\\
(8)& if the lock level of a partition is larger than zero, there should be a process in this partition disabled the preemption \\
(9)& if there is a process that disabled the preemption of this partition, the lock level of the partition should be larger than zero\\
(10)& if the lock level of a partition is zero, the partition should be in the $NORMAL$ mode  \\

(11)& if the current process and current partition are valid, the process should be in the partition  \\
(12)& the validation of current partition implies that the partition's mode is not $IDLE$\\
(13)& the validation of current process implies that the process is running and its partition is in $NORMAL$\\
(14)& if a process was delayed started, it has a delay time \\
(15)& the aperiodic process has the special (infinite) value of period  \\
(16)& the periodic process has a finite value of period \\
& \\
&\textbf{Inter- / Intra- partition communication} \\
(17)& the message queue size of the queuing port is finite \\
(18)& the message number in the queue of queuing port is not larger than the maximum number of messages \\
(19)& the message queue size of the buffer is finite \\
(20)& the message number in the queue of buffer is not larger than the maximum number of messages\\
(21)& if the empty indicator of a blackboard is $OCCUPIED$, there must be a message on it\\
(22)& the value of the semaphore is not larger then the maximum value \\
(23)& if a process is waiting for a buffer, the partition that it belongs to is also the partition that the buffer belongs to. And the same as the blackboard, semaphore, and event \\
(24)& if a process is in the waiting queue of a queuing port, the process state should be $Waiting$. And the same as the buffer, blackboard, semaphore, and event \\
& \\
&\textbf{Health monitoring} \\
(25)& the error handler has maximum priority  \\
(26)& the error handler is in its partition where the handler was created\\
(27)& the error handler and the process which created the handler are in the same partition \\

\bottomrule
\end{tabular}
\label{tbl:invariants}
\end{table}
\section{Formalizing ARINC 653}
\label{sec:form}

This section presents a formalization of ARINC 653 in Event-B. We first discuss our formalization criteria, then the Event-B model of the key system functionalities\footnote{The complete Event-B model can be downloaded from GitHub, https://github.com/ywzh/arinc653model}, and finally how service requirements are formalized by translating the service specification grammar into Event-B.

\subsection{Formalization criteria}
The criteria of modeling Event-B based systems are how to represent the system components, their attributes, safety properties, and actions by  Event-B components such as \emph{constants}, \emph{sets}, \emph{variables}, and \emph{events}. 

\textbf{Components and types}. The main components in ARINC 653 are: \emph{partition}, \emph{process}, \emph{communicating components} (port, channel, buffer, blackboard, semaphore, and event), \emph{waiting queue}, and \emph{error handler}. These components can be classified as statically configured components and dynamic components. Partitions are static components configured at build time and initialized during the POS booting. Any other component is a dynamic component, which is only created during partition initialization, using services ARINC 653 \emph{CREATE\_*}. In particular, ARINC 653 does not allow creating components during partition run-time. We use \emph{sets} in Event-B to represent partitions. Process and communicating components are represented by \emph{sets} and  \emph{variables}. Sets specify the valid domain of a component, whilst variables are used to keep track of the created components during initialization. For instance, set $PROCESSES$ defines the domain of all processes in a system and variable $processes$ stores already created processes, and is a subset of $PROCESSES$. Waiting queues and error handlers are associated with communicating components and partitions respectively, and are represented as \emph{variables}.

\textbf{Component attributes}. ARINC 653 defines the attributes of each component including fixed attributes and variable attributes. Fixed attributes can not be changed during run-time and are defined as \emph{constants} in Event-B, and variable attributes are defined as \emph{variables}. These constants and variables are functions mapping from a component's set to the data type of the attribute. Event-B supports functions with different properties, such as \emph{partial functions}, \emph{total functions}, and \emph{partial injections}. For instance, the $Period\_of\_Partition$ constant is a partition attribute, and represents a total function $Period\_of\_Partition \in  PARTITIONS \tfun  \nat$, to indicate that all partitions have a \emph{period} attribute of type natural. Whilst, the $deadlinetime\_of\_process$ process attribute is a partial function $deadlinetime\_of\_process \in  processes \pfun  \nat$, since deadline time is undefined when a process is stopped, and defined when the process is in any other state.

\textbf{Component relations}. Due to spatial separation of ARINC 653, each component is created in one partition and it is bound to that partition. These relations may be one-to-one or one-to-many, which are represented with different types of functions in Event-B. For instance, $errorhandler\_of\_partition \in  PARTITIONS \pinj  processes$ is a partial injection since each partition has at most one error handler process. 

\textbf{Control and actions}. ARINC 653 defines control and actions on components, such as the partition control, the process control, and the port control. Each action on these components are formalized with one or more \emph{events} in Event-B. Due to the semantic gap between sequential description of service requirements and the \textit{guard-action} style event model in Event-B, we carefully considered the design principle of events and used semi-automatic translation from service requirements to Event-B as discussed in {\subsectprefix~\ref{subsec:trans}}.

\subsection{Event-B model of key system functionality}
This subsection introduces the Event-B model for some of the key system functionalities in ARINC 653. 

\subsubsection{Partition and process}

Partitions and processes functionality mainly considers partition control (operating modes and transition), process control (process states and transitions), and the scheduling principle. 
Since time management considers the timing aspect on control of process execution, we also model this functionality in this part. 



Process state transitions are complex since process states are dependent on partition operating modes. 
Process states and transitions are encoded in Event-B in machine \evtbmach{Mach\_ PartProc\_ Trans} as it follows. From the guard of the $process\_state\_transition$ event ($\mathbf{grd20}$ and $\mathbf{grd21}$), we can analyze the nesting between partition operating modes and process states. This event models most of the state transitions, however some state transitions are modelled separately in other events because they are a consequence of other systems events. For instance, transiting from state $Ready$ to state $Running$ is triggered by the process scheduling, therefore this transition is modelled in event $scheduling$; the state transitions from process states in the \arinctext{COLD\_START} or \arinctext{WARM\_START} partition modes to states in the \arinctext{NORMAL} mode are modelled in event \evtbmach{partition\_modetransition\_to\_normal}; the creation of a process is modelled in event $create\_process$, hence the transition from state $Dormant$ to $Ready$ is carried out in that event. In ARINC 653, state $Waiting$ implies three situations, a process was \emph{suspended}, \emph{waiting for a resource}, or \emph{suspended when waiting for a resource}. To explicitly distinguish these situations, we define three states to represent them $Suspend$, $Waiting$ and $WaitandSuspend$ respectively.

\noindent
\fbox{\parbox[t]{3.4in}{
\scriptsize 
\textbf {process\_state\_transition} $\defi$ \textbf{any} $part \; proc \; newstate$ \textbf{where} \\
\mbox{  } \parbox{0.35in}{\textbf{@grd01 }} \parbox{2.8in}{$ part \in  PARTITIONS $;} \\
\mbox{  } \parbox{0.35in}{\textbf{@grd02 }} \parbox{2.8in}{$ proc \in  processes $;} \\
\mbox{  } \parbox{0.35in}{\textbf{@grd03 }} \parbox{2.8in}{$ newstate \in  PROCESS\_STATES $;} \\
\mbox{  } \parbox{0.35in}{\textbf{@grd06 }} \parbox{2.8in}{$ processes\_of\_partition(proc) = part $;} \\
\mbox{  } \parbox{0.35in}{\textbf{@grd07 }} \parbox{2.8in}{$ partition\_mode(part) \neq  PM\_IDLE $;} \\
\mbox{  } \parbox{0.35in}{\textbf{@grd20 }} \parbox{2.8in}{$ ((partition\_mode(part) = PM\_COLD\_START  \lor  \\ partition\_mode(part) 
			    = PM\_WARM\_START)  \land \\ process\_state(proc) = PS\_Dormant) 
			    \limp \\ newstate = PS\_Waiting $;} \\
\mbox{  }  \parbox{0.35in}{\textbf{@grd21 }} \parbox{2.8in}{$ ((partition\_mode(part) = PM\_COLD\_START \lor \\ partition\_mode(part) 
			   = PM\_WARM\_START) \land  \\ process\_state(proc) = PS\_Waiting) 
			   \limp \\  (newstate = PS\_Dormant \lor \\ newstate = PS\_WaitandSuspend) $;} \\
\mbox{  } \parbox{0.35in} ...... //here we omit other detailed guards \\
		\textbf{then} \\
\mbox{  } \parbox{0.35in}{\textbf{@act01 }} \parbox{2.8in}{$ process\_state(proc) :=  newstate $;} \\
		\textbf{end}
}}	

Process state transitions in Event-B only model the possible transition path, not the actions executed during the transitions. 
These are modelled in machine \evtbmach{Mach\_PartProc\_Trans\_withEvents}, which models the process control, and defines the actions triggered by state transitions by means of the events $suspend$, $suspend\_self$, $resume$, $stop$, $stop\_self$, $start$, $delayed\_start$, $timed\_wait$, $period\_wait$, $time\_out$, $req\_busy\_resouce$, $resource\_become\_available$, among others, for the concrete process control. In these events, we strengthen the guards of $process\_state\_transition$ and those events defined in machine \evtbmach{Mach\_PartProc\_Trans\_withEvents}, in such a way that it is refined by \evtbmach{Mach\_PartProc\_Trans\_withEvents}.

\subsubsection{Timing and scheduling}

Timing and scheduling  functionalities are important features of safety-critical real-time systems. Timing is not explicitly specified in ARINC 653 as a system functionality, except time management for processes. We provide a simple model of timing and a two-level scheduling, to schedule IMA partitions and processes.


\textbf{Scheduling}.
In POSs, all actions are initiated whenever a significant event occurs but scheduling, which is initiated by the regular event of a clock tick. 
The tick-tock has been considered as a way to model discrete time in Event-B \cite{moha11}. We use  event $ticktock$ to represent the regular event of a clock tick, which increases variable $clock\_tick$ in one unit and variable $need\_reschedule$ is set to $TRUE$ when appropriate to trigger the scheduling. 
The scheduling specified in ARINC 653 is a two-level scheduling. Partition scheduling is a fixed, cycle based scheduling and is strictly deterministic over time. The cyclic scheduling consists of a major time frame (\emph{MTF}) that is split into partition time windows (\emph{PTW}), each \emph{PTW} having starting and ending time relatives to the starting time of the \emph{MTF}. Each \emph{PTW} of a \emph{MTF} is associated to a given partition, which is executed when the system time reaches the starting point of the \emph{PTW} and a new partition scheduling is performed when the system time reaches the end of the \emph{PTW}. Process scheduling is priority preemptive and carried out by the partition. 

The scheduling process chooses the current partition and process under execution according to the current \emph{PTW} and the process priority. 
The partition scheduling is enabled when variable $need\_reschedule$ is $TRUE$, and the current time is in a \emph{PTW} different than the current one. Note that the \arinctext{IDLE} partition can not be scheduled. Variable $need\_reschedule$ is enabled by event $ticktock$ or by the events in which the running process is blocked. We also define variable $need\_procresch$ to indicate process scheduling after partition scheduling. If the partition mode is \arinctext{NORMAL}, an ARINC process should be chosen to be executed. Otherwise, the partition is at the \arinctext{START} mode and the main process of this partition occupies the processor time. 
There are two special types of ARINC processes that need to be carefully considered during process scheduling: the error handler and processes locking the process preemption of its partition. 
A process can lock the process preemption of its partition to prevent process rescheduling when accessing a critical section or a resource shared by multiple processes of the same partition. The error handler is a special aperiodic process with the highest priority, without deadline, and is invoked by the POS when a process level error is detected. It preempts any running process regardless of its priority and even if preemption is locked.

The abstract meaning of process scheduling for a partition in the \arinctext{NORMAL} mode is to choose a process to run. The current running process is placed into the $Ready$ state and the chosen process into the $Running$ state. If the chosen process is the current running process, it remains in the $Running$ state. The abstract event of process scheduling is shown as it follows. 

The $process\_schedule$ abstract event is extended in machine \evtbmach{Mach\_PartProc\_Manage} by events $process\_schedule$ and $run\_errorhandler\_preempter$. The first event represents the normal process scheduling in a partition, whilst the second one represents the situation where the error handler of the partition has been started or a process has locked the preemption process  of this partition. 

\noindent
\fbox{\parbox{3.4in}{
\scriptsize 
\textbf {process\_schedule} $\defi$ \textbf{any} $part \; proc$ \textbf{where} \\
\mbox{  } \parbox{0.35in}{\textbf{ @grd01 }} \parbox{2.8in}{$ part \in  PARTITIONS $;}\\
\mbox{  } \parbox{0.35in}{\textbf{ @grd02 }} \parbox{2.8in}{$ proc \in  processes $;}\\
\mbox{  } \parbox{0.35in}{\textbf{ @grd03 }} \parbox{2.8in}{$ processes\_of\_partition(proc) = part $;}\\
\mbox{  } \parbox{0.35in}{\textbf{ @grd04 }} \parbox{2.8in}{$ partition\_mode(part) = PM\_NORMAL $;}\\
\mbox{  } \parbox{0.35in}{\textbf{ @grd05 }} \parbox{2.8in}{$ process\_state(proc) = PS\_Ready \lor \\ process\_state(proc) = PS\_Running$;}\\
\textbf{then} \\
\mbox{  } \parbox{0.35in}{\textbf{ @act1 }} \parbox{2.8in}{$ process\_state :=  (process\_state \ovl  \\ (process\_state^{-1} [\{ PS\_Running\} ] \cprod  \{ PS\_Ready\} )) \ovl \\ \{ proc \mapsto   PS\_Running\}  $;}\\
\textbf{end}
}}

\begin{figure}
\centering
\includegraphics[width=3.0in]{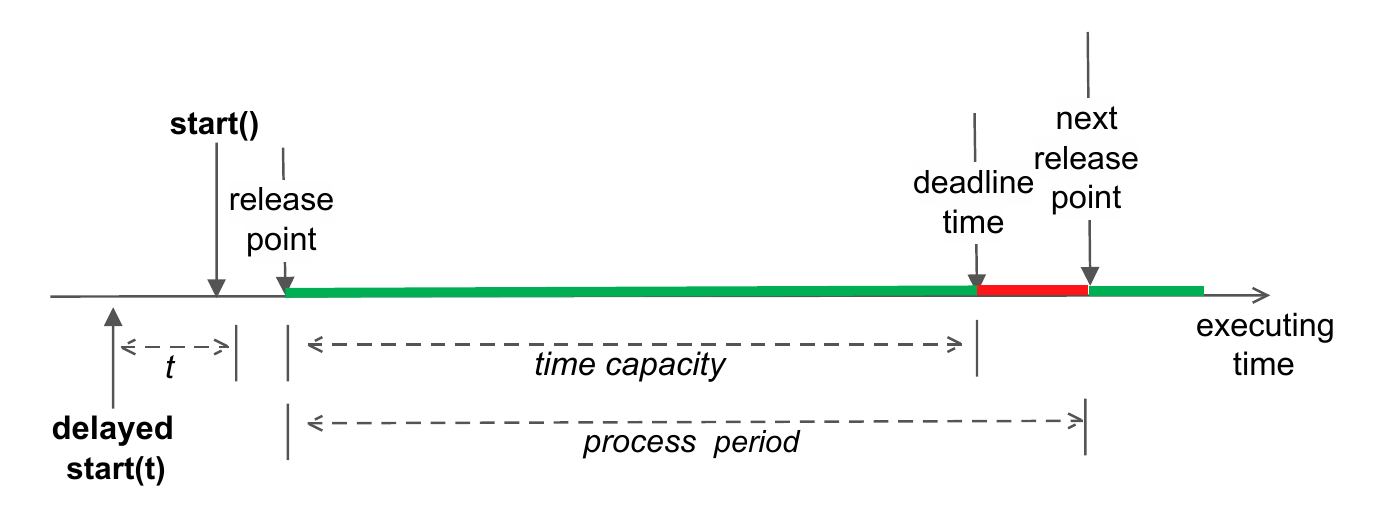}
\caption{Timing model of the periodic process} 
\label{fig:periodic_proc}
\vspace{-10pt}
\end{figure}

\textbf{Timing model for processes}. 
ARINC 653 distinguishes between periodic and aperiodic processes for the reason that they have different timing models. 
Whilst timing model for aperiodic processes is relatively simple, 
the timing model for periodic processes is more complex, as shown in {\figprefix} \ref{fig:periodic_proc}. Periodic processes have a release point time, a time capacity, a period, and a deadline time. When a periodic process is started, it is placed into the \emph{Waiting} state to wait for its release point, which is the first periodic process start in the next \emph{MTF}. The process deadline time is its release point plus its time capacity, i.e. the maximum allowed time for that process to be under execution. When a process reaches its release point, its next release point is calculated as the current release point plus its period. If a periodic process is started with a delay $t$, its release point and deadline time are calculated based on the starting time plus the delay $t$. 

Periodic processes can also be \emph{timed waited} for a delay time $t$. The \emph{timed waited} process is placed into the \emph{Waiting} state and waken up after time $t$, remaining unchanged the deadline time of the process. The \emph{periodic wait} event suspends the execution of a periodic process until its next release point, so the new release point is its current release point plus the process period. Similarly, the deadline time of the process is also recalculated based on the new release point. The \emph{replenish} event updates the deadline time of a periodic process with a specified \emph{budget time}. If a periodic process finishes before the \emph{deadline time}, it is placed into the \emph{Waiting} state to wait for the next release point and the POS asks for process rescheduling. Otherwise, its deadline time is missed and an exception is arisen that is handled by the health monitor. Since the deadline time of a process is in absolute time, it is easy to determine a deadline time miss when the current time exceeds the deadline time: $clock\_tick * ONE\_TICK\_TIME >  deadlinetime\_of\_process(proc)$. When a periodic process waiting for its next release point reaches it, event $periodicproc\_reach\_releasepoint$ is triggered and new release point and deadline times are recalculated. Note that \emph{Suspend} and \emph{resume} can not be carried out on periodic processes.

\textbf{Time-out trigger}.
Another timing functionality in ARINC 653 is the \emph{time-out trigger}. When a process is delayed started, is suspended by itself, or sends a message to a buffer with a \emph{time-out}, the POS initiates a time counter for that time-out. After the time elapses, the POS responds to the expiration of the time-out by starting the process or sending a time-out message. 

We define variable $timeout\_trigger \in processes \pfun  (PROCESS\_STATES \cprod  \nat_1)$ to store the time counter for processes. $(proc \mapsto (PS\_Ready \mapsto t)) \in timeout\_trigger$ means that a process $proc$ is blocked waiting for some resource until time $t$, moment at which the blocked process is placed into state $Ready$. 
In the model, a tuple $(ps \times t)$ is inserted into $timeout\_trigger$ when the POS initiates a time counter with a duration $t$, and it blocks the process. Then when $t$ arrives, the event $time\_out$ triggers a blocked process to state $ps$.


\subsubsection{Interpartition communication}
Interpartition communication is conducted via messages in ARINC 653. A partition is allowed to exchange messages through multiple channels via their respective source and destination ports. Ports can be configured into two different modes which determine the communication mode: queueing and sampling modes. Interpartion communication related actions and events are modelled in the machine \evtbmach{Mach\_IPC\_Conds}.

\begin{figure}
\begin{center}
\includegraphics[width=2.5in]{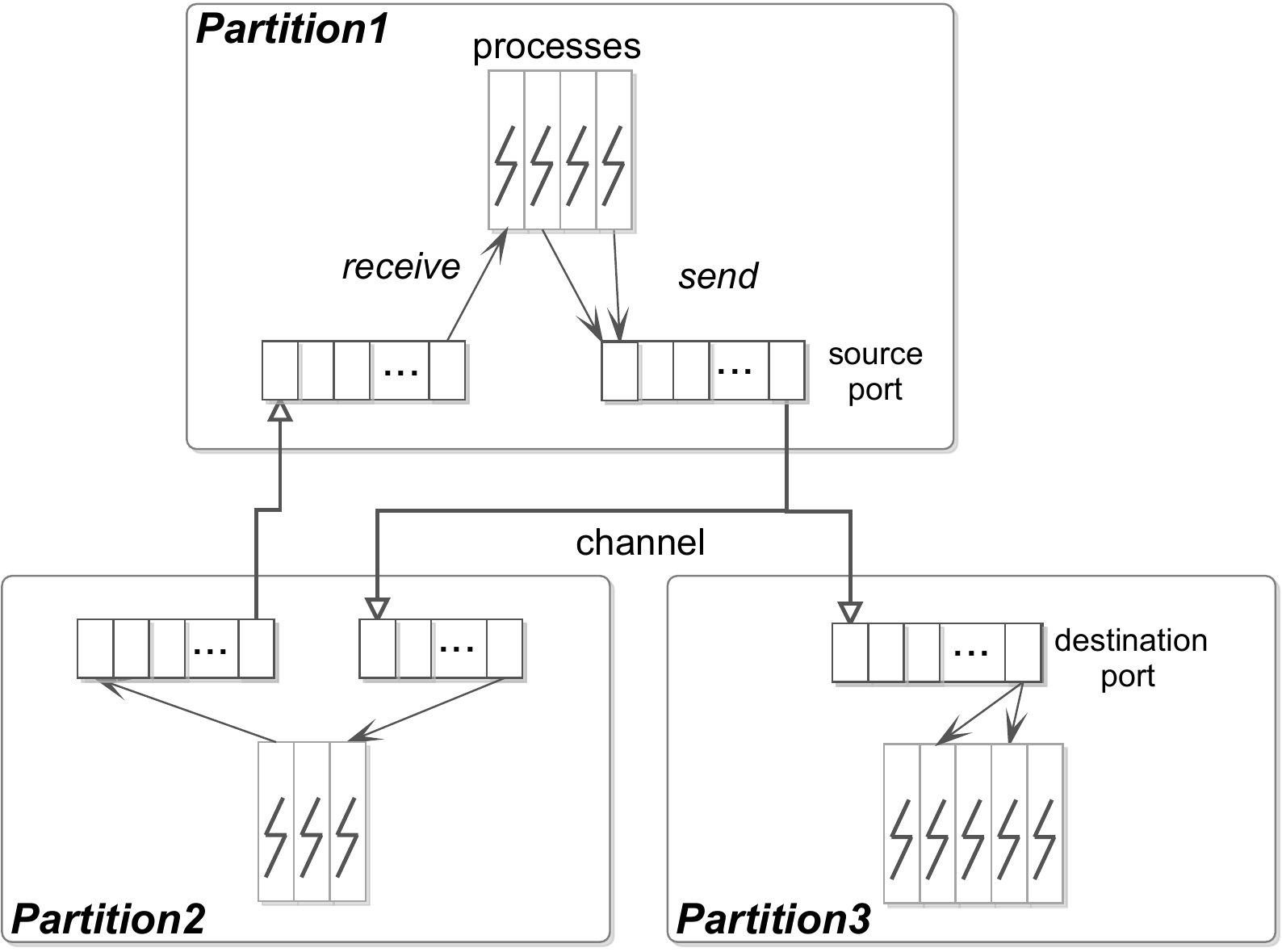}
\caption{The message based interpartition communication (queuing mode)} 
\label{fig:interpartcomm}
\end{center}
\vspace{-10pt}
\end{figure}

Messages are atomic entities in ARINC 653. For this reason, we define set $MESSAGES$ to represent all abstract messages in the system, and variable $used\_messages$ to represent the set of already sent messages (used). In interpartition communication sending services, messages to be sent should be in set $MESSAGES \setminus used\_messages$, and stored into $used\_messages$ after being sent. 

Each port has a message space to store the message(s) to be sent/received via this port. On the one hand, sampling ports have a single message storage, which is modelled using variable $msgspace\_of\_samplingports\in SamplingPorts \pfun   (MESSAGES\cprod \nat_1)$. On the other, queueing ports keep a message queue, which is represented by variable $queue\_of\_queueingports\in QueuingPorts\tfun \pow(MESSAGES\cprod \nat_1)$. A pair in $MESSAGES\cprod \nat_1$ means a message sent/received at a specific time. For instance, if the service of writing sampling message $m$ has been invoked on port $p$ at time $t$, then the ordered pair $p \mapsto (m \mapsto t) \in msgspace\_of\_samplingports$. 

``The communication between partitions is done by processes which are sending or receiving messages. In queuing mode, processes may wait on a full message queue (sending direction) or on an empty message queue (receiving direction). Processes waiting for a port in queuing mode are queued in FIFO or priority order.''~\cite{ARINC653p1} We use variable $processes\_waitingfor\_queuingports\in (processes\cprod \nat_1\cprod MESSAGES)\pfun QueuingPorts$ to store waiting processes of a queuing port. An ordered pair $(proc \mapsto t \mapsto m) \mapsto port \in processes\_waitingfor\_queuingports$ means that process $proc$ invoked the service of sending/receiving a message $m$ to/from $port$ on the time $t$ and was blocked. ARINC 653 specifies two different policies to unblock processes blocked in a queuing port: based on the highest waiting time, where the unblocked process is that one which has been blocked in the queue for a longer time, or based on the process priority, where the unblocked process is that one with a higher priority among those blocked in the queuing port. This policy is statically specified for each queuing port. The message based communication in the queuing mode among partitions is shown in {\figprefix} \ref{fig:interpartcomm}.

The port control system functionality in interpartition communication describes the communicating actions including creating ports, and sending/receiving message(s) to/from a sampling or queuing port. These actions are specified as events indicating the guards and the result of the port control action, which is proven to be strengthened and simulated by machine $MACH\_IPC$, which encodes the service requirements, hence refining machine  $MACH\_IPC\_Conds$. The event $send\_queuing\_message$ is modelled as it follows.

\noindent
\fbox{\parbox{3.4in}{
\scriptsize 
\textbf {send\_queuing\_message} $\defi$ \textbf{any} $port \; msg$ \textbf{where} \\
\mbox{  } \parbox{0.35in}{\textbf{ @grd01 }} \parbox{2.8in}{$ part \in  PARTITIONS $;}\\
\mbox{  } \parbox{0.35in}{\textbf{ @grd01 }} \parbox{2.8in}{$ port\in ports $;}\\
\mbox{  } \parbox{0.35in}{\textbf{ @grd02 }} \parbox{2.8in}{$ port\in QueuingPorts $;}\\
\mbox{  } \parbox{0.35in}{\textbf{ @grd03 }} \parbox{2.8in}{$Direction\_of\_Ports(port)=PORT\_SOURCE $;}\\
\mbox{  } \parbox{0.35in}{\textbf{ @grd04 }} \parbox{2.8in}{$ msg\in MESSAGES \land  msg\notin used\_messages $;}\\
\mbox{  } \parbox{0.35in}{\textbf{ @grd05 }} \parbox{2.8in}{$ card(queue\_of\_queueingports(port))< \\ MaxMsgNum\_of\_QueuingPorts(port) $;}\\
\mbox{  } \parbox{0.35in}{\textbf{ @grd06 }} \parbox{2.8in}{$ processes\_waitingfor\_queuingports^{-1} [\{ port\} ] = \emptyset  $;}\\
\textbf{then} \\
\mbox{  } \parbox{0.35in}{\textbf{ @act01 }} \parbox{2.8in}{$ queue\_of\_queueingports:| \exists t\qdot (t\in \nat \land  (msg\mapsto t) \in \\ queue\_of\_queueingports'(port)) $;}\\
\mbox{  } \parbox{0.35in}{\textbf{ @act02 }} \parbox{2.8in}{$ used\_messages :=  used\_messages \bunion  \{ msg\}  $;}\\
\textbf{end}
}}

This event encodes the action of sending a message $msg$ via a queuing port $port$ when the port is not full and there is no other process waiting to send a message through this port. Guard $\mathbf{grd05}$ ensures that the current number of messages of the queue is less than the queue size. Guard $\mathbf{grd06}$ ensures that there is not any other process waiting for sending messages via this port. Otherwise, the process invoking the event needs to be blocked. The actions state that message $msg$ is in the queue of $port$ ($\mathbf{act01}$) and has been used ($\mathbf{act02}$) after the execution of this event.

\subsubsection{Intrapartition communication}

Intrapartition communication mechanisms are buffers, blackboards, semaphores, and events. For space reasons we only introduce the \emph{blackboard} model.

A blackboard allows to send/receive messages to/from processes belonging to the same partition. Any message written on it remains there until the message is either cleared or overwritten by a new instance of the message. When a process attempts to read a message from an empty blackboard, it is queued for a specified amount of time. When a message is displayed on the blackboard, the POS removes from the process queue all the processes waiting for that blackboard and puts them in state $Ready$. Variables $msgspace\_of\_blackboards\in blackboards \pfun  MESSAGES$ and $emptyindicator\_of\_blackboards \in  blackboards \tfun  BLACKBOARD\_INDICATORTYPE$ store the message that has been displayed on a blackboard and  respectively indicate whether a blackboard is empty or not. If $emptyindicator\_of\_blackboards$ is not false then $msgspace\_of\_blackboards$ is different from the empty set, therefore indicating that the blackboard contains at least one message. Events $display\_blackboard$ and $display\_blackboard\_needwakeuprdprocs$ encode the functionality of displaying a message in the blackboard when there are not any waiting process for that blackboard and when there are any waiting process that have to be awaken. Similarly, events $read\_blackboard$ and $read\_$ $blackboard\_whenempty$ respectively encode the reading functionality when the blackboard is not empty or it is empty and the process has to be enqueued in the waiting process queue. Event $display\_blackboard\_needwakeuprdprocs$ extends the event $resource\_become\_available2$ that wakes up all processes waiting for a resource, and is enabled when the waiting process queue is not empty ($\mathbf{grd505}$). Actions $\mathbf{act501} \sim \mathbf{act504}$ represent the result of this event. The Event-B model is shown below.

\noindent
\fbox{\parbox{3.4in}{
\scriptsize 
\textbf {display\_blackboard\_needwakeuprdprocs} \textbf{extends} \textbf {resource\_become\_available2} $\defi$ \textbf{any} $bb \; msg$ \textbf{where} \\
\mbox{  } \parbox{0.37in}{\textbf{ @grd500 }} \parbox{2.8in}{$ bb\in blackboards $;}\\
\mbox{  } \parbox{0.37in}{\textbf{ @grd504 }} \parbox{2.8in}{$ msg\in MESSAGES \land  msg\notin used\_messages $;}\\
\mbox{  } \parbox{0.37in}{\textbf{ @grd505 }} \parbox{2.8in}{$ processes\_waitingfor\_blackboards^{-1} [\{ bb\} ] \neq  \emptyset  $;}\\
\textbf{then} \\
\mbox{  } \parbox{0.37in}{\textbf{ @act501 }} \parbox{2.8in}{$ msgspace\_of\_blackboards(bb) :=  msg $;}\\
\mbox{  } \parbox{0.37in}{\textbf{ @act502 }} \parbox{2.8in}{$ processes\_waitingfor\_blackboards :=  procs\domsub \\ processes\_waitingfor\_blackboards $;}\\
\mbox{  } \parbox{0.37in}{\textbf{ @act503 }} \parbox{2.8in}{$ used\_messages :=  used\_messages \bunion  \{ msg\}  $;}\\
\mbox{  } \parbox{0.37in}{\textbf{ @act504 }} \parbox{2.8in}{$ emptyindicator\_of\_blackboards(bb) :=  \\ BB\_OCCUPIED $;}\\
\textbf{end}
}}

\subsubsection{Health monitoring}

In ARINC 653, HM is responsible for responding to and reporting hardware, application and POS software faults and failures. ARINC 653 supports HM by providing HM configuration tables and an application level error handler process. These tables are the Module HM table, the Multi-Partition HM tables, and the Partition HM tables. The error handler is a special process of the partition with the highest priority and no process identifier.

The HM is modelled in machine \evtbmach{Mach\_HM}. It defines variable $module\_shutdown\in BOOL$ to control the module, and each event in  machine \evtbmach{Mach\_IPC} are extended in \evtbmach{Mach\_HM} by adding a guard $module\_shutdown = FALSE$. This means that if the module is shut down, no event can be triggered. 

The HM decision logic is implemented in the \emph{guards} of the events encoding recovery actions. 
Recovery actions at the module level are only triggered when the error is inside a partition time window (\emph{PTW}) and the error is a module level error of the partition executed during the current \emph{PTW}. 
Recovery actions at the partition level are triggered when the error is a partition level error and the error handler has not been created in that partition, or the error was caused by the error handler of the partition. 
For instance, the event $hm\_recoveryaction\_shutdown\_module$ is specified as it follows.

\noindent
\fbox{\parbox{3.4in}{
\scriptsize 
\textbf {hm\_recoveryaction\_shutdown\_module} $\defi$ \textbf{any} $errcode$ \textbf{where} \\
\mbox{  } \parbox{0.37in}{\textbf{ @grd700 }} \parbox{2.8in}{$ module\_shutdown = FALSE $;}\\
\mbox{  } \parbox{0.37in}{\textbf{ @grd701 }} \parbox{2.8in}{$ errcode\in SYSTEM\_ERRORS $;}\\
\mbox{  } \parbox{0.37in}{\textbf{ @grd702 }} \parbox{2.8in}{$ errcode\in dom(MultiPart\_HM\_Table(part)) $;}\\
\mbox{  } \parbox{0.37in}{\textbf{ @grd703 }} \parbox{2.8in}{$ errcode \mapsto  MLA\_SHUTDOWN \in \\ MultiPart\_HM\_Table(part) $;}\\
\textbf{then} \\
\mbox{  } \parbox{0.37in}{\textbf{ @act701 }} \parbox{2.8in}{$ module\_shutdown:= TRUE $;}\\
\textbf{end}
}}

%

Finally, if the error level is process, the error handler of this partition has been created, and the error was not caused by the error handler, the error handler is activated to deal with this error. How to handle an error is application dependent. Event $hm\_recoveryaction\_errorhandler$ is specified as it follows. The guard $\mathbf{grd703}$ means that a \emph{Deadline\_Missed} error occurs when some process in this partition missed its deadline time. 

\noindent
\fbox{\parbox{3.4in}{
\scriptsize 
\textbf {hm\_recoveryaction\_errorhandler} \textbf{extends} \textbf {start\_aperiodprocess\_innormal} $\defi$ \textbf{any} $errcode$ \textbf{where} \\
\mbox{  } \parbox{0.37in}{\textbf{ @grd700 }} \parbox{2.8in}{$ module\_shutdown = FALSE $;}\\
\mbox{  } \parbox{0.37in}{\textbf{ @grd701 }} \parbox{2.8in}{$ errcode\in SYSTEM\_ERRORS $;}\\
\mbox{  } \parbox{0.37in}{\textbf{ @grd702 }} \parbox{2.8in}{$ (errcode\in dom(Partition\_HM\_Table(part)) \land \\ \exists a\qdot (a\in PARTITION\_RECOVERY\_ACTIONS \land \\ ERROR\_LEVEL\_PROCESS\mapsto a\in \\ dom(Partition\_HM\_Table(part)(errcode)))  ) $;}\\
\mbox{  } \parbox{0.37in}{\textbf{ @grd703 }} \parbox{2.8in}{$ DEADLINE\_MISSED\in ran(Partition\_HM\_Table\\(part)(errcode)) \limp (\exists proc\qdot (proc\in \\ processes\_of\_partition^{-1} [\{ part\} ] \land \\ clock\_tick* ONE\_TICK\_TIME > \\ deadlinetime\_of\_process(proc))) $;}\\
\mbox{  } \parbox{0.37in}{\textbf{ @grd704 }} \parbox{2.8in}{$ part\in dom(errorhandler\_of\_partition) $;}\\
\mbox{  } \parbox{0.37in}{\textbf{ @grd705 }} \parbox{2.8in}{$ current\_process \neq  errorhandler\_of\_partition(part) $;}\\
\mbox{  } \parbox{0.37in}{\textbf{ @grd706 }} \parbox{2.8in}{$ proc = errorhandler\_of\_partition(part) $;}\\
\textbf{end}
}}

\subsection{Translating service requirements into Event-B}
\label{subsec:trans}

 A service definition in the APEX specification contains the name of the service and a list of formal parameters ($p_1$...$p_n$) with their types. The description of a service consists of a short definition of its functionality and a full description of its semantics. The semantic description gives the algorithm of the service behavior, which is composed of two parts: an error part which describes error handling due to incorrect values of actual input or input-output parameters, and a normal part which describes the treatment to be performed when no error is detected by the service.

\begin{figure}
\begin{center}
\includegraphics[width=3.0in]{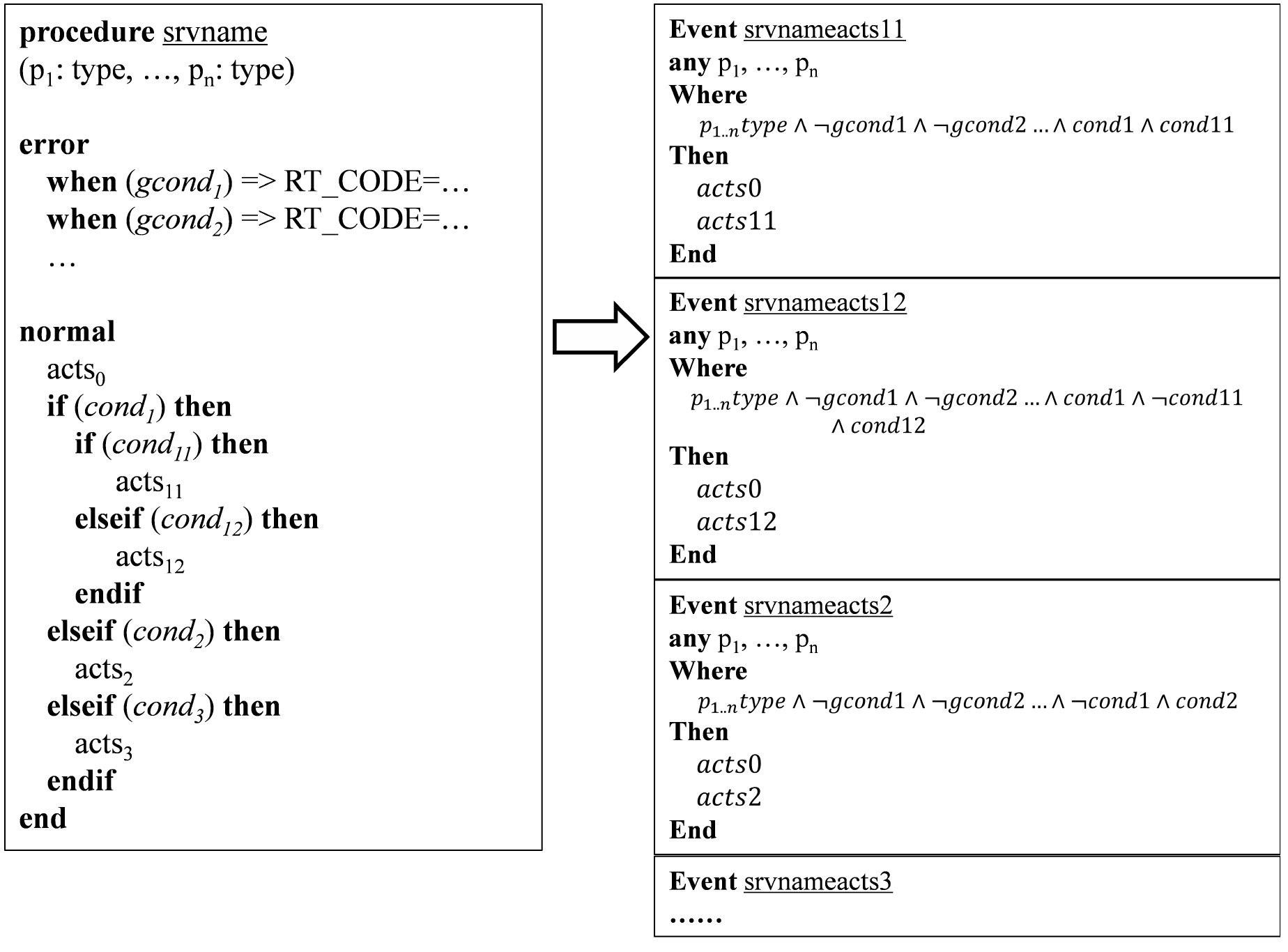}
\caption{Translate APEX specification into Event-B} 
\label{fig:apex_trans}
\end{center}
\vspace{-10pt}
\end{figure}

Although ARINC 653 defines a structured language to describe the functional requirements of APEX services, we find that ARINC 653 Part 1 only uses compound statements ``IF'' and ``SEQUENCE''  to describe complex structures. 
Moreover, due to the natural language description of simple statements, it is difficult to implement automatic translation from APEX specification to Event-B model. Hence, we concentrate on translating the structure of  APEX services to events, and the detailed behavior of each service represented by simple statements needs to be modelled by hand. The specification grammar for APEX services is illustrated in the left part of {\figprefix} \ref{fig:apex_trans}, and the generated events are shown in the right part.

Event-B does not have compound statements such as ``IF'', ``CASE'', and ``LOOP'', therefore the description of an APEX service should be decomposed into events with non-intersect guards. That is, if one event is enabled by its guard, then all of other events are disabled. For the ``IF'' statement, its body (e.g., $acts_{11}$, $acts_{12}$, $acts_2$, $acts_3$) are behaviors under different conditions that do not intersect. Therefore, we use an event to represent the behavior of each body. Additionally, for each ``IF'' statement we need to add a new event representing the ``ELSE'' body, even when it is and ``IF'' without an ``ELSE''.
The type of parameters of the APEX service is encoded as guards of each event ($p_{1..n}type$). 
In the error part, conditions $gcond_1$, $gcond_2$, etc. indicate incorrect values of parameters, and thus their negation are translated into Event-B as \emph{guards} of each event. The conditions of the ``IF'' statement are also translated as guards. In Fig~\ref{fig:apex_trans}, the condition of $acts_{11}$ is $cond_1 \wedge cond_{11}$, so the guard of the corresponding event includes $cond_1 \wedge cond_{11}$. Whilst, the condition of $acts_{12}$ is $cond_1 \wedge \neg cond_{11} \wedge cond_{12}$. 

A simple statement, such as ``set the specified process state to DORMANT; '' in {\figprefix} \ref{fig:stop_service}, is translated into actions in the event according to the meaning of the statement, which is usually represented by a deterministic assignment, e.g., $process\_state(proc) :=  PS\_DORMANT$.

We have designed an algorithm to guide manual translations from APEX service requirements into Event-B models as shown in {\algprefix} \ref{alg:trans}. For convenience, we first give a simple syntax for the APEX service specification grammar:
\begin{equation*}
\scriptsize
\begin{aligned}
c\; ::= \; & \mathbf{ACT} \; act \;| \;c;c \; | \mathbf{IF}\; cond \;\mathbf{THEN}\; c\; | &\mathbf{IF} \; cond \; \mathbf{THEN}\;c\;\mathbf{ELSE}\;c
\end{aligned}
\end{equation*}
where $\mathbf{ACT}\; act$ is a simple statement and $c;c$ is the sequence statement.

\begin{algorithm}
\scriptsize
\textbf{function} translate($evts$,$stmt$)\{\\
~~\Switch{$stmt$}{
    \Case{$\mathbf{ACT}$ act}{
        \emph{add the action $act$ to end of action list of each event in $evts$;}\\
        \Return $evts$\;
    }
    \Case{st1;st2}{
        $evts' \leftarrow translate(evts,st1)$;\\
        \Return $translate(evts', st2)$;
    }
    \Case{$\mathbf{IF}$ cond $\mathbf{THEN}$ st1 $\mathbf{ELSE}$ st2}{
        $evts' \leftarrow \; duplicate \; of \; evts$;\\
        \emph{add the ``$cond$'' to end of guard list of each event in $evts$};\\
        $evts \leftarrow translate(evts,st1)$;\\
        \emph{add the ``$\neg cond$'' to  end of guard list of each event in $evts'$};\\
        $evts' \leftarrow translate(evts',st2)$;\\
        \Return $evts \cup evts'$
    }
    \Case{$\mathbf{IF}$ cond $\mathbf{THEN}$ st}{
        $evts' \leftarrow \; duplicate \; of \; evts$;\\
        \emph{add the ``$cond$'' to end of guard list of each event in $evts$};\\
        $evts \leftarrow translate(evts,st1)$;\\
        \emph{add the ``$\neg cond$'' to end of guard list of each event in $evts'$};\\
        \Return $evts \cup evts'$
    }
}
\}\\
\textbf{function} translate\_service($spec$)\{  //$spec = <\zeta, P, E, S>$\\
$\;\;$ $evts \leftarrow \{<\zeta,\phi,\phi>\}$\\
$\;\;$ $evts \leftarrow translate(evts,S)$\\
$\;\;$ \emph{add $\wedge_i(\neg E_i)$ to guard list of each event in $evts$} \\
$\;\;$ \Return \{ev. ev$\in evts \wedge$ $ev.\alpha \neq  \emptyset$\}\\
\}
\caption{Translate APEX service into Event-B}\label{alg:trans}
\end{algorithm}

The algorithm translates a service requirement presented in this syntax into a set of events. A service requirement is a tuple $<\zeta, P, E, S>$, where $\zeta$ is the service name, $P$ is a parameter list, $E$ is the error conditions in the error part, and $S$ is the normal part. In {\algprefix} \ref{alg:trans}, function $translate$ translates a statement into a set of events and $translate\_service$ translates a service requirement into the final set of events. An event in the $evts$ set is a tuple $<\iota,\sigma,\alpha>$, where $\iota$ is name of the event, $\sigma$ is a list of guards, and $\alpha$ is the list of actions of the event. We initially put an event with empty guards and empty actions into set $evts$. The actions for simple statements is immediate, but the case of statement ``IF'' is not so straightforward. For each ``IF'' it is necessary to create two events: one having as guards the ``IF'' condition, and the other its negation. Then, their bodies are recursively processed by the translation algorithm, adding actions to the events when the body is a simple action, or creating new events with non-intersect guards in case the body contains a nested ``IF''. The composition statement ``;'' is trivial in the case of two simple statements or a simple statement and an ``IF'' statement. However, the case of two  ``IF'' statements is slightly more elaborated. In this case, the first ``IF'' creates a set of events $evts$ that are passed as an argument to the translation of the second ``IF''. Then the second ``IF'' duplicates $evts$ into $evts'$, and adds its conditions as guards to each event in $evts$. Similarly, it adds the negated condition to each event in $evts'$, therefore obtaining all possible non-intersecting guards for both ``IF''. Finally, the algorithm adds the negative of the conditions in the error part to the guard list of each event, and it removes those events with empty actions. The event name in the final event set is manually renamed according to their meanings.


Finally, this translation approach can be simplified when the bodies of ``IF'' statements are very simple, e.g., when the body is only a simple statement. In that case the conditions of an ``IF'' statement can be represented in the guard of the event. For instance, in the specification of the STOP service ({\figprefix} \ref{fig:stop_service}),  the first ``IF'' (``current process is error handler and PROCESS\_ID is the process which the error handler preempted''), the partition's lock level is reset. In the event $stop$, we add a new parameter $newlocklevel$ and two guards shown as it follows.

\noindent
\fbox{\parbox{3.4in}{
\scriptsize 
\parbox{0.35in}{\textbf{ @grd45 }} \parbox{2.8in}{$ current\_process\_flag = TRUE \land \\ part\in dom(errorhandler\_of\_partition) \land \\ current\_process = errorhandler\_of\_partition(part) \land \\ proc = process\_call\_errorhandler(current\_process) \\ \limp  newlocklevel = \{ part \mapsto  0\}  $;}\\
\parbox{0.35in}{\textbf{ @grd46 }} \parbox{2.8in}{$ \lnot (current\_process\_flag = TRUE \land \\ part\in dom(errorhandler\_of\_partition) \land \\ current\_process = errorhandler\_of\_partition(part) \land \\ proc = process\_call\_errorhandler(current\_process)) \\ \limp  newlocklevel = \emptyset $;}\\
}}

$\mathbf{grd45}$ means that if the current process is the error handler of the current partition and the process to be stopped is the process which the error handler preempted, then $newlocklevel$ is an ordered pair $ part \mapsto  0$. Otherwise, $newlocklevel$ is $\emptyset$ ($\mathbf{grd46}$). In the actions of event $stop$, there is an assignment $locklevel\_of\_partition :=   locklevel\_of\_partition \ovl newlocklevel$. Thus, the semantic of the ``IF'' statement is implemented without decomposing into different events.

\section{Results and discussion}
\label{sec:result}

\subsection{Model and proof statistics}
In {\tabprefix} \ref{tbl:statistics} we give model and proof statistics of ARINC 653 Part 1 in the Rodin tool \footnote{The LOC in this table is counted in the original Event-B editor of Rodin. While for pretty printing, the Event-B model that can be downloaded from our web page is printed from Camille editor which is a plugin of Rodin}. These statistics measure the size of the model, the proof obligations generated and discharged by the Rodin tool, and those proved interactively. The LOC of the machines increase gradually since the refinement machine is an extension of the refined machine. Therefore, the total LOC is not the summation of all machines and is not counted here. 

\begin{table}
\centering
\scriptsize
\caption{Statistics of model and proof} 
\begin{tabular} {L{3cm}R{0.5cm}R{1.1cm}R{1.2cm}R{1.2cm}}
\toprule
Machine & LOC & Proof obligations& Automatically discharged & Interactively discharged  \\
\midrule
Mach\_Part\_Trans&39&7&6 (86\%)&1(14\%)\\
Mach\_PartProc\_Trans&236&128&118 (92\%)&10(8\%)\\
Mach\_PartProc\_Trans\_withEvents&495&223&219 (98\%)&4(2\%)\\
Mach\_PartProc\_Manage&900&580&504 (87\%)&76(13\%)\\
Mach\_IPC\_Conds&2022&272&169 (62\%)&103(38\%)\\
Mach\_IPC&2267&560&463 (83\%)&97(17\%)\\
Mach\_HM&2712&18&9 (50\%)&9(50\%)\\
\midrule
Total &&1791 & 1491(83\%) &300(17\%)  \\
\bottomrule
\end{tabular}
\label{tbl:statistics}
\end{table}

Proving is a time-consuming and skilled work in Event-B. Fortunately, the Rodin tool provides effective automatic provers that saves much proving time on our model. In addition it is possible to integrate third-party provers (such as Atelier B prover) and SMT solvers (such as CVC3 and Z3) as Rodin plugings bringing the degree of automation to a higher level (more than 80\%). 

\subsection{Errors found in ARINC 653}
We found three errors in ARINC 653 Part 1, one in the process management service, and two in the inter- and intra-partition communication services. Additionally we detected three cases where the specification of process state transitions is incomplete.

\subsubsection{In process state transitions}
An incomplete description of process state transitions is detected. The ``process states and state transitions'' description in ARINC 653 Part 1 is very detailed and try to list all actions triggering transitions. But we find that some significant actions are not involved either in the figure or in the text. 

The errors are the following:
\begin{deflist}
\item in the \arinctext{COLD/WARM\_START} mode, a suspended process can also be resumed and its state transits from \emph{Waiting} to \emph{Waiting}. This action is missed in the ``Waiting - Waiting'' transition conditions in the standard. 
\item in the \arinctext{NORMAL} mode, if an aperiodic process is \emph{delayed\_started} (if the delay time $>$ 0), its state transits from \emph{Dormant} to \emph{Waiting}. This action is missed in the ``Dormant - Waiting'' transition conditions in the NORMAL mode in the standard.
\item in the \arinctext{NORMAL} mode, if an aperiodic process is \emph{delayed\_started} (i.e., the delay time = 0), its state transits from \emph{Dormant} to \emph{Ready}. This action is missed in the ``Dormant - Ready'' transition conditions in the \arinctext{NORMAL} mode in the standard.
\end{deflist}

These incompleteness are found by verifying the \emph{guard strengthening} between machine \evtbmach{Mach\_PartProc\_} \evtbmach{Trans\_withEvents} and its refinement machine \evtbmach{Mach\_PartProc\_Manage}. The first machine models the process state transitions and their triggering actions according to the standard, and the refinement models the service requirements. The guard strengthening requires that the state transitions specified in the refinement should not be contradictory with the transitions in \evtbmach{Mach\_PartProc\_Trans\_withEvents}. After carefully checking these errors, we find that the service requirements are correct while the process state transitions are incomplete.






\subsubsection{In requirement of process management services}
%

The error is in the requirements of the \emph{RESUME} service. In fact, an aperiodic process that has been \emph{delayed\_started} is in the \emph{Waiting} state if it is \emph{suspended}. When that process is resumed, it should retain in the \emph{Waiting} state if the delay time has not been reached. 
But according to service requirement \emph{RESUME}  defined in ARINC 653 partly shown below, the aperiodic process is set into the \emph{Ready} state, because this process is not waiting on a process queue or on a TIMED\_WAIT time delay. This error is found by verifying the \emph{guard strengthening} between the $resume$ events in the machine \evtbmach{Mach\_PartProc\_Trans\_with\_Events} and \evtbmach{Mach\_PartProc\_Manage}.

\begin{center}
\begin{lstlisting}[frame=single, breaklines=true, basicstyle=\tiny, numberstyle=\scriptsize] %\tiny,\scriptsize,\footnotesize,\small,

if (the specified process is not waiting on a process queue or TIMED_WAIT time delay) then 
    set the specified process state to READY; 
    if (preemption is enabled) then 
        ask for process scheduling; 
        -- The current process may be preemptedby the resumed process
    end if; 
end if; 
\end{lstlisting}
\end{center}






\subsubsection{In requirements of the communication service}
We find two errors in the requirements of the communication service. 

The first one is an error in inter-partition communication. The $SEND\_QUEUING\_MESSAGE$ service  is used to send a message via a specified queuing port. If there is enough space in the queuing port to accept the message, the message is inserted to the end of the port's message queue. If there is not, then the process is blocked and stays in the waiting queue until the specified time-out, if finite, expires, or space becomes free in the port to accept the message. However, in the service specification when the time-out does not expire and space is released, the sent message is not inserted into the message queue.  This leads to an error where sending processes waiting for the free space within the time-out loses their sent messages. 
This error is found by verifying the \emph{simulation} between the $send\_queueing\_message$ events in the machine \evtbmach{Mach\_PartProc\_Trans\_with\_Events} and \evtbmach{Mach\_PartProc\_Manage}. The event in the first machine requires that the message should be in the message queue after being sent when the message queue is sufficient, or not sufficient but the sending process does not expires. But the second machine is contradictory with it.

The second error is in the specification of the \emph{RECEIVE\_BUFFER} service in intra-partition communication. This service is used to receive a message from a specified buffer. When the buffer is not empty, the receiving process can receive a message directly, and the message should be removed from the message queue of this buffer. But in the service specification, we find that the received message is not removed. This leads to an error that the message queue of a buffer may always be full. This error is found by verifying the \emph{simulation} between the $receive\_message$ events in machines \evtbmach{Mach\_PartProc\_Trans\_with\_Events} and \evtbmach{Mach\_PartProc\_Manage}. The event in the first machine requires that the message should not be in the message queue after being received, but the second machine is contradictory with it.





\subsection{Discussion}

\subsubsection{Completeness of our model}
Although we have formalized all of ARINC 653 services and the system functionality, some implementation-related details are not covered in our formalization: (1) we eliminate execution context attributes of partitions and processes, e.g.  entry point and stack size, since  ARINC 653 only defines these attributes, but do not provide any functionality for them. (2) we omit the detailed partition switch, that leads us to capture only the error inside a partition time window in the HM. (3) we do not model the booting/initialization of the POS, since ARINC 653 specifies nothing about the POS booting process. For the purpose of verifying the ARINC standard or ARINC based applications, these eliminations do not affect the verification result. However, these details should be implemented when formally developing a POS from the ARINC 653 specification. 

\subsubsection{Event-B modeling of software specification}
The most notable problem when we are formalizing the ARINC 653 is the semantic gap between sequential description of the service requirements and \emph{guard-action} style event model. ARINC 653 specifies the software, i.e. the partitioning operating system, therefore the semantics of each service are specified by a structural sequential language, whilst Event-B is a formalism for developing and verifying systems using \emph{events}. Sequential programs can be generated from the Event-B model, while to our knowledge, there is no known work on how to translate structural languages into Event-B. This paper provides a preliminary translation approach from APEX specification grammar to Event-B, but an ideal and semantically equivalent translation framework is needed for high assurance.


\subsubsection{Deductive verification vs. model checking}
Model checking is an automatic and ``push-down'' verification approach for system and software. As concluded in \cite{Beck14}, verification of complex systems is never automatic or ``push-button''. This standpoint is also confirmed in our work. In the Rodin tool, the ProB \cite{leusc03} is a model checker for the Event-B models. At first, we try to furnish properties in linear temporal logic and model check them using ProB. We find that it is feasible on our model on the first and second levels of abstraction, but the state space is too large to be checked on other levels. 
The deductive verification in Event-B is done by logic reasoning. Although it needs expert skills in mathematics and logics, most of proof obligations can be discharged automatically by provers (up to 80\%). 

\subsubsection{Safety and liveness properties}
The shortcoming of deductive verification in Event-B is that properties are represented as \emph{invariants}, which are safety properties. Liveness properties are another major class of properties \cite{Schn87}. In \cite{Dwyer99}, the authors collected 555 examples of property specifications, 27.2\% are invariant properties (global absence and universality properties), whilst, liveness properties cover about 45.6\%(global existence and response).
We found that liveness properties are very useful in the specification of the communication in ARINC 653. 
Liveness properties can partially be expressed by the Flow language in \cite{Alex11}, but it is difficult to specify possible continuation scenarios, which means that \emph{globally} ($G$) quantifier is not supported. 
Reasoning about liveness properties directly on Event-B model is feasible \cite{Hoang11}, the absence of supporting tools makes it difficult to be applied to complex models that lead us to forgo the verification. 




\subsubsection{Reusing models for development and verification}
The Event-B model of ARINC 653 defines the abstract specification of POSs. This model can also be used in the model-based development and the verification of POSs. 

A POS is an execution environment for applications. Composition of the ARINC 653 and application models enables the simulation, analysis, and verification for IMA systems. This requires that the application is also modelled in Event-B and that tools in Event-B support complex analysis on the model. Current version of the Rodin tool only supports functional verification and simulation. In particular, time analysis needs to be strengthened in Rodin or else it is necessary to export the Event-B model to another analysis tools. 

Second, the ARINC 653 model provides the possibility of formally developing a new POS by refinement, and finally generating the source code. The Rodin tool provides many code generation plugins, such as C, C++, Java and Ada. The difficulty of this goal may come from the  hardware dependency of POSs and the efficiency of the generated code. The source code of POSs usually has some intrinsic patterns and contains assembly code. These require that the Event-B model is detailed enough and that the code generation in Rodin is revised according to this target. 





\section{Conclusions}
\label{sec:concl}
In this work, we have presented the formalization and deductive verification of the ARINC 653 standard using Event-B. The system functionality and all of 57 services specified in ARINC 653 Part 1 have been modelled in Event-B. The safety properties of the system functionality and service requirements, and the consistency between them are checked by discharging proof obligations of invariants and refinement preservation. Finally, we found three errors in ARINC 653 Part 1 and detected three cases where the specification is incomplete. The verification was significantly simplified due to the high degree of automated reasoning in the Rodin tool.



As future work we consider to include mechanical checking of these errors in ARINC 653 compliant OSs source code such as XtratuM, and POK, and to extract liveness properties from the ARINC 653 standard and discover appropriate solutions to verify them. Since ARINC 653 is being considered to support multicore platform, it is also an important aspect in our future work. 


\section*{Acknowledgement}
We much appreciate the suggestions from Prof. Jean-Paul Bodeveix and Prof. Mamoun Filali from IRIT, Universit\'e de Toulouse, France. 
This work is partially supported by the Fundamental Research Funds for the Central Universities in China (Grant No.YWF-15-GJSYS-083), and by the National Research Foundation, Prime Minister’s Office, Singapore under its National Cybersecurity R\&D Program (Award No. NRF2014NCR-NCR001-30) and administered by the National  Cybersecurity R\&D Directorate.

\bibliographystyle{IEEEtran}
\bibliography{verify_arinc653}

\end{document}